# Intraspecific Comparative Genomics of *Candida albicans* Mitochondria Reveals Non-Coding Regions Under Neutral Evolution

Thais F. Bartelli[1], Renata C. Ferreira[1,2], Arnaldo L. Colombo[2], Marcelo R. S. Briones[1, *]

[1]Laboratório de Genômica Evolutiva e Biocomplexidade, Departamento de Microbiologia, Imunologia e Parasitologia, Disciplina de Microbiologia, Universidade Federal de São Paulo, Rua Pedro de Toledo, 669, 4º andar, fundos, São Paulo, SP, CEP 04039-032, Brazil. [2]Laboratório Especial de Micologia, Disciplina de Infectologia, Universidade Federal de São Paulo, Rua Pedro de Toledo, 669, 5º andar, CEP 04039-032, São Paulo, SP, Brazil.

*** Corresponding author:**
Marcelo R. S. Briones
Laboratório de Genômica Evolutiva e Biocomplexidade, Departamento de Microbiologia, Imunologia e Parasitologia, Universidade Federal de São Paulo, Rua Pedro de Toledo, 669, 4º andar, fundos, São Paulo, SP, CEP 04039-032, Brazil.
Phone: 5511 5576-4537 Fax: 5511 5572-4711 E-mail:marcelo.briones@unifesp.br

**Abstract**

The opportunistic fungal pathogen *Candida albicans* causes serious hematogenic hospital acquired candidiasis with worldwide impact on public health. Because of its importance as a nosocomial etiologic agent, *C. albicans* genome has been largely studied to identify intraspecific variation and several typing methods have been developed to distinguish closely related strains. Mitochondrial DNA can be useful for this purpose because, as compared to nuclear DNA, its higher mutational load and evolutionary rate readily reveals microvariants. Accordingly, we sequenced and assembled, with 8 fold coverage, the mitochondrial genomes of two *C. albicans* clinical isolates (L296 and L757) and compared these sequences with the genome sequence of reference strain SC5314. The genome alignment of 33,928 positions revealed 372 polymorphic sites being 230 in coding and 142 in non-coding regions. Three intergenic regions located between genes tRNAGly*/COX1*, *NAD3/COB* and ssurRNA*/NAD4L*, named IG1*,* IG2 and IG3 respectively, which showed high number of neutral substitutions, were amplified and sequenced from 18 clinical isolates from different locations in Latin America and 2 ATCC standard *C. albicans* strains. High variability of sequence and size were observed, ranging up to 56bp size difference and phylogenies based on IG1, IG2 and IG3 revealed three groups. Insertions of up to 49bp were observed exclusively in Argentinean strains relative to the other sequences which could suggest clustering by geographical polymorphism. Because of neutral evolution, high variability, easy isolation by PCR and full length sequencing these mitochondrial intergenic regions can contribute with a novel perspective in molecular studies of *C. albicans* isolates, complementing well established multilocus sequence typing methods.

## 1. Introduction

*Candida* spp. are important opportunistic fungal pathogens and one of the major leading causes of superficial and life-threatening bloodstream infections, especially in hospitalized immunocompromised hosts (Koh et al. 2008; Lim et al. 2012; Pfaller,1996). In Brazil, the overall incidence reported in a surveillance study showed 2.49 cases per 1,000 hospital admissions which is 2 to 15 times greater than in countries in the Northern Hemisphere, such as the United States (Colombo et al. 2006). The primary source of most of these infections is endogenous, though there is severe risk of acquisition of *Candida* spp. from the hospital environment by contaminated plastic devices and staff skin (Dorko et al. 1999; Fanello et al. 2001; Pfaller, 1996).

The genome of *C. albicans* has been extensively studied to identify intraspecific variability and several typing methods were developed to effectively elucidate the epidemiology of *C. albicans* and to discriminate clinical isolates to help identify the source of contamination (Cliff et al. 2008; Fanello et al. 2001). DNA fingerprinting methods such as restriction fragment length polymorphism (RFLP), randomly amplified polymorphic DNA (RAPD) and pulsed field gel electrophoresis (PFGE), have been widely used for *C. albicans* typing (Fanello et al. 2001; Heo et al. 2011; Noumi et al. 2009; Ruiz-Diez et al. 1997). However, these techniques are prone to ambiguity and subjective interpretations because of variations in electrophoretic patterns such as band size and intensity. Moreover, these techniques are not indicated for estimating genetic distances and phylogenetic inference, because they underestimate the real number of evolutionary events, are subject to systematic errors and cannot be readily assessed in terms of probability models (Mello et al. 1998; Soll, 2000). More reliable molecular studying methods based on sequencing, such as the gold standard multilocus sequence typing (MLST), relies on the analysis of at least six nuclear housekeeping genes (Robles et al. 2004) and though several authors have used *C. albicans* mtDNA in molecular analysis (Anderson et al. 2001; Aranishi, 2006; Jacobsen et al. 2008; Sanson and Briones, 2000; Watanabe et al. 2005), more studies are needed to investigate fully its intraspecific nucleotide diversity in *C. albicans*.

Mitochondrial DNA (mtDNA) is more susceptible to damage and mutations than nuclear DNA, mainly because of the presence of reactive oxygen species generated during ATP synthesis and less efficient repair system of gamma DNA polymerase (Kang and Hamasaki, 2002; Kaguni, 2004). The high mutation number and the faster evolutionary rate, from 5 to 10 times higher than nuclear DNA (Brown et al. 1979), makes mtDNA suitable for discrimination of closely related organisms and recent evolutionary events. Furthermore, because it is haploid and present in multiple copies in cells, greater efforts and high technology are not usually required for the amplification and sequencing of specific PCR products. Despite the high variability of mitochondrial genes, their use can be limited in genetic analysis of closely related populations because of low intraspecific variability, probably constrained by negative selection on functional domains (Aranishi, 2006; Sanson and Briones, 2000; Watanabe et al. 2005). Non-coding regions (*e.g.* introns, pseudogenes, intergenic) evolve neutrally or are at least significantly less susceptible to natural selection and fitness interference than coding regions. Therefore, these genomic segments are expected to have a higher number of polymorphic sites and to evolve faster, making them interesting sequences to explore intraspecific mitochondrial nucleotide variability (Aranishi, 2006; Watanabe et al. 2005).

In this study, we have sequenced the complete mitochondrial genomes of two *C. albicans* clinical isolates and compared them with the genome sequence of the reference strain SC5314, to identify intraspecific hypervariable sites. We demonstrated that intergenic regions evolves under neutrality and are the most variable segments in the mtDNA, interesting features that could bring light into the usefulness of these sequences in molecular studies of *C. albicans* microvariability.

## 2. Materials and methods

### 2.1. Strains and mtDNA isolation.

*C. albicans* clinical isolates were obtained from the collection of the “Laboratório Especial de Micologia (LEMI), Disciplina de Doenças Infecciosas e Parasitárias (DIPA), Departamento de Medicina, Universidade Federal de São Paulo”. 18 isolates were collected from patients with hematogenic infection by *C. albicans* from 1997 to 2010 in different locations in Latin America. Two standard *C. albicans* ATCC (American Type Culture Collection) strains were also used in the analysis (Table 1). Cultures were grown in YPD medium (1% yeast extract, 2% peptone, 2% dextrose) at 30ºC before experiments. Mitochondrial DNA for whole genome sequencing or PCR amplifications was isolated by the method described previously (Defontaine et al. 1990).

**Table 1.** *C. albicans* clinical isolates and accession number of nucleotide sequences used in this study. IG1=tRNA-Gly/*COX1,* IG2=*NAD3/COB and* IG3=ssurRNA/*NAD4L*. Brazilian states RJ (Rio de Janeiro); SP (São Paulo); PR (Paraná); BA (Bahia). Strains in bold indicate that the complete mitochondrial genome sequence was used as source for nucleotide sequence. *COB*=Cytochrome *b*, ITS1 and ITS2=rDNA ITS excluding 5.8S rDNA.

| | Source | | | GenBank accession no. | | | | | |
|---|---|---|---|---|---|---|---|---|---|
| Strain | Clinical | Geographic | Year | IG1 | IG2 | IG3 | *COB* | ITS1 | ITS2 |
| **SC5314** | Blood | USA | | NC002653 | NC002653 | NC002653 | NC002653 | NC002653 | NC002653 |
| ATCC 24433 | Nail | USA | | JQ814087 | JQ814119 | JQ814140 | - | JX494812 | JX494813 |
| ATCC 90029 | Blood | USA | | JQ814086 | JQ814120 | JQ814141 | - | JX494814 | JX494815 |
| 34 ptc | Catheter | ? | ? | JQ814102 | JQ814105 | JQ814125 | - | JX494790 | JX494791 |
| **L296** | Blood | Brazil / RJ | 1997 | JQ864234 | JQ864234 | JQ864234 | JQ864234 | JQ814076 | JQ814082 |
| **L757** | Blood | Brazil / SP | 2001 | JQ864233 | JQ864233 | JQ864233 | JQ864233 | JQ814077 | JQ814083 |
| 6965 | Blood | Brazil / SP | 2010 | JQ814098 | JQ814109 | JQ814129 | - | JX494798 | JX494799 |
| 6944A | Blood | Brazil / SP | 2010 | JQ814100 | JQ814106 | JQ814127 | - | JX494794 | JX494795 |
| 7060A | Blood | Brazil / SP | 2010 | JQ814097 | JQ814123 | JQ814130 | - | JX494800 | JX494800 |
| 6945 | Blood | Brazil / SP | 2010 | JQ814099 | JQ814108 | JQ814128 | - | JX494796 | JX494797 |
| 6921 | Blood | Brazil / PR | 2010 | JQ814101 | JQ814107 | JQ814126 | - | JX494792 | JX494792 |
| 7252A | Blood | Brazil / PR | 2010 | JQ814094 | JQ814112 | JQ814133 | - | JX494804 | JX494805 |
| 7251 | Blood | Brazil / PR | 2010 | JQ814095 | JQ814111 | JQ814132 | JQ814068 | JQ814072 | JQ814078 |
| 7082 | Blood | Brazil / PR | 2010 | JQ814096 | JQ814110 | JQ814131 | - | JX494802 | JX494803 |
| 6924 | Blood | Brazil / BA | 2010 | JQ814103 | JQ814104 | JQ814124 | - | JX494788 | JX494789 |
| 5147 | Blood | Ecuador | 2009 | JQ814089 | JQ814117 | JQ814138 | - | JQ814074 | JQ814080 |
| 6592 | Blood | Ecuador | 2009 | JQ814091 | JQ814115 | JQ814136 | - | JX494808 | JX494809 |
| 5982 | Blood | Argentina | 2009 | JQ814088 | JQ814118 | JQ814139 | JQ814067 | JQ814075 | JQ814081 |
| 6779 | Blood | Argentina | 2009 | JQ814090 | JQ814116 | JQ814137 | JQ814069 | JX494810 | JX494811 |
| 6185 | Blood | Venezuela | 2009 | JQ814093 | JQ814113 | JQ814134 | - | JX494806 | JX494807 |
| 6461 | Blood | Colombia | 2009 | JQ814092 | JQ814114 | JQ814135 | - | JQ814073 | JQ814079 |

## 2.2. Yeast nuclei purification and DNA extraction.

Yeast nuclei purification was performed according to the method described previously (Hahn, 2006). Nuclear DNA was extracted by adding 200 µl of Solution B (100 mM NaCl, 10 mM EDTA, 1% Sarkosyl, 50 mM Tris-HCl pH 7.8) and incubated for 30 min at room temperature, followed by purification with phenol-chloroform, washed in 70% Ethanol, Ethanol precipitated and ressuspended in TE buffer.

## 2.3. Whole mitochondrial genome sequencing and assembly.

The complete mitochondrial genome sequences of two *C. albicans* clinical isolates (L296 and L757) were obtained using the whole genome shotgun method (Fleischmann et al. 1995). For mitochondrial genomic library construction, mtDNA was randomly sheared by sonication (Sambrook and Russel, 2001) and fragments of size from 1 to 2kb were blunt cloned into pBluescript IISK (Stratagene) prior to sequencing. mtDNA sequences were determined by dideoxynucleotide chain termination method of Sanger et al. (1977) using fluorescent BigDye terminator cycle sequencing kit (version 3.1; Applied Biosystems) in an ABI Prism 3100 automated sequencer (Applied Biosystems) according to the manufacturer's instructions. Assembly of finished sequences from chromatograms was generated using Phred (Ewing and Green, 1998; Ewing et al. 1998a), Phrap and Consed (Gordon et al. 1998). Sequences were considered finished when Phred scores were above 40, which corresponds to less than one estimated error per 10 kb assembled.

## 2.4. Amplification of mitochondrial intergenic regions.

PCR primers were designed for complete amplification of nucleotide sequences of three *C. albicans* mitochondrial intergenic regions according to the available sequence of the reference strain SC5314 (GenBank ID: NC002653.1) (Table 2). Amplification reactions (50 µl) consisted of 10 mM dNTP, 10 pmol of each primer (forward and reverse), 10 µl Buffer B (2 mM $MgCl_2$), 40 ng mtDNA and 1µl Elongase Enzyme Mix (Invitrogen). For the mitochondrial intergenic region located between the genes tRNA-Gly/*COX1* (IG1) cycling conditions were 94ºC for 5 min, followed by 35 cycles of 94ºC for 40 s, 48ºC for 40 s, 68 ºC for 1 min and a final extension step of 68ºC for 7 min; for *NAD3*/*COB* (IG2) conditions were 94ºC for 5 min, 35 cycles of 94ºC for 45 s, 50ºC for 45 s, 68ºC for 1 min and extension of 68ºC for 7 min while PCR cycling conditions for the sequence flanked by ssurRNA/*NAD4L* (IG3) were 94ºC for 5 min, followed by 35 cycles of 94ºC for 45 s, 48ºC for 45s, 68ºC for 2 min and extension of 68ºC for 7 min. Amplicons were blunt cloned into pBluescript II SK (Stratagene) before sequencing. Sequencing reactions were performed as described previously in the section 2.3. PCR products were also sequenced on both strands by using the same primers employed in the amplification.

**2.5. *Cytochrome b* gene (*COB*) and rDNA ITS (internal transcribed spacer) amplification.**

PCR primers (A and L) were used for complete amplification of *COB* gene (Table 2). Amplification reactions (50 µl) consisted of 10 mM dNTP, 10 pmol of each primer, 10 µl Buffer B (2 mM MgCl2), 40 ng mtDNA and 1µl Elongase Enzyme Mix (Invitrogen). Cycling conditions were 94ºC for 5 min, followed by 35 cycles of 94º for 40 s, 50ºC for 40 s, 68ºC for 4 min and final extension of 68ºC for 7 min. Total genomic DNA was extracted as described previously (Wach et al. 1994) and ITS amplification was performed using universal primers ITS1 and ITS4 (Table 2) (White et al. 1990). Amplification reactions (25 µl) consisted of 12.5 µl of 2X Master Mix (Fermentas), 10 pmol of each specific primer (forward and reverse) and 40 ng of DNA with the following cycling conditions: 94ºC for 5 min, 35 cycles of 94ºC for 30 s, 58ºC for 30 s, 72ºC for 50 min and final extension of 72ºC for 7 min. *COB* and ITS PCR products were sequenced on both strands as described in section 2.3 using the same corresponding primers. For the complete sequencing of *COB* (2,811 bp), specific internal primers (Primers COB B to COB K - Table 2) were also used.

**2.6. Comparative sequence analysis.**

Alignment of the whole mitochondrial genomes was made using Geneious 4.8 (Drummond et al. 2012) by the progressive Mauve algorithm (Darling et al. 2004). Nucleotide sequences of mitochondrial intergenic regions, *COB* and ITS, were aligned using Clustal W (Thompson et al. 1994). The overall pairwise mean distances (p-value) of the intergenic regions, *COB* and ITS were estimated using the program MEGA 5 (Tamura et al. 2011) with pairwise deletion treatment of gaps.

**Table 2.** Primers used for *C. albicans* DNA amplification and sequencing.

| Primer | Gene or region amplified | Sequence 5' > 3' | Amplicon size (bp) |
|---|---|---|---|
| RIG 1 forward | tRNA-Gly/*COX1* | GCCAGGGTCTACCATTA | 635 |
| RIG 1 reverse | (IG1) | CATAGCACTAACCATACC | |
| RIG 2 forward | *NAD3*/*COB* | GCGTAGTTATGATAAGGATA | 836 |
| RIG 2 reverse | (IG2) | GTATTAGATTTACGTGTTGGC | |
| RIG 3 forward | ssurRNA/*NAD4L* | GCTATAAGTTGAAATACAGT | 1188 |
| RIG 3 reverse | (IG3) | AGTAATGTAGTAATAACAGC | |
| COB A | *COB* | GTAGTGGAGGTGCTTATATAC | 3109 |
| COB L | | GAGCTATAGTTCACTTACC | |
| COB B | | CTATTGTAAGAAGTGTTACC | |
| COB C | | CATGCTAATGGTGCCTCA | |
| COB D | | CTTTAGGACTATCCGCTTG | |
| COB E | | GAGGTAGTAAACCATTAAAG | |
| COB F | | CCGGTCAATCTTTATTTCC | |
| COB G | | CGTAGTATAGAGAAAGGTT | |
| COB H | | CCACGGTCTTGATTTAGTC | |
| COB I | | GGCAAATGAGTCATTGAGG | |
| COB J | | CAATTGAAGGAGGTGTTAC | |
| COB K | | GCCAATGCATCCTTACTTC | |
| ITS 1 | rDNA ITS | TCCGTAGGTGAACCTGCGG | 536 |
| ITS 4 | | TCCTCCGCTTATTGATATGC | |
| ACT 1 forward | *ACT1* | GAAGCTCCAATGAATCCAAAATC | 355 |
| ACT 1 reverse | | GTTCGAAATCCAAAGCAACGTAAC | |
| COX 2 forward | *COX2* | ATGCGAGGTATATCGGTTC | 947 |
| COX 2 reverse | | GCGATTCCACTAATTAAGG | |

## 2.7. Phylogenetic inference and testing for neutral evolution.

Phylogenetic trees were generated by the Bayesian method using the program MrBayes (Huelsenbeck and Ronquist, 2001). Trees were inferred from $10^6$ generations sampling a tree in every 100 generations until the standard deviation from split frequencies were under 0.01. The parameters and the trees were summarized by wasting 25% of the samples obtained (burnin). The consensus trees (50%) were then used to determine the posterior probabilities values. Substitution models were optimized by ModelTest 3.7 (Posada and Crandall, 1998). All phylogenetic trees were then formatted with the FigTree v1.3.1 program (http://tree.bio.ed.ac.uk/software/figtree/). Statistical tests of Tajima's D (Tajima, 1989) and Fu and Li's (Fu and Li, 1993) D* and F* for detection of deviation from the neutral model of evolution were performed using DnaSP 5 (Librado and Rozas, 2009).

## 2.8. *Actin* and *COX2* amplification reactions.

After nuclei isolation and DNA extraction, a fragment of approximately 350 bp from the nuclear gene *ACT1* (positions 274 to 628) and the complete sequence of the mitochondrial gene *COX2* (Table 2), employed as a positive and negative control respectively, were amplified by PCR. For *ACT1*, amplification reaction (25 µl) consisted of 12.5 µl of 2X Master Mix (Fermentas), 10 pmol of each specific primer (forward and reverse) and 40 ng of DNA. Cycling conditions were 94ºC for 5 min, 35 cycles of 94ºC for 40 s, 54ºC for 40 s, 72ºC for 1 min and final extension of 72ºC for 10 min. For *COX2,* reaction and cycling were the same as described for *ACT1,* except that the primer annealing temperature used was 50ºC. Amplification of mitochondrial intergenic regions with the purified nuclei DNA sample was performed according to the protocol described elsewhere in section 2.4.

## 2.9. Nucleotide sequences accession number.

*C. albicans* sequences obtained in this study have been deposited in GenBank (http://www.ncbi.nlm.nih.gov/nucleotide/) under the accession numbers listed in Table 1.

## 3. Results

### 3.1. Intraspecific comparative sequence analysis of *C. albicans* mitochondrial genome.

The complete mitochondrial genomes of two *C. albicans* clinical isolates (L296 and L757) were sequenced by the whole genome shotgun approach, with 8-fold coverage. The final mtDNA assemblies were 33,928 bp (assembly error: 0.2/10kb) and 33,631 bp (assembly error: 0.01/10kb) for strains L757 and L296 respectively. Genome annotation was performed using program ORF finder as implemented in Geneious 4.8 (Drummond et al. 2012) and was consistent with the annotation of reference strain SC5314 mitochondrial genome (assembly 19, available online in the *Candida* genome database - candidagenome.org), as confirmed by BLAST (Basic Local Alignment Search Tool). To avoid redundancy in alignments the two identical repeat regions of 6,842 bp present in the *C. albicans* SC5314 (40,420 bp) mtDNA, were represented only once in the final assembly of strains L296 and L757.

Alignment of mtDNA from strains L296, L757 and SC5314 revealed 372 polymorphic sites in the 33,928 nucleotide sites analyzed, corresponding roughly to 1.1% global variation (Fig.1), where 230 (0.68 %) of these polymorphisms are in coding regions and 58.70% are transitions (Table 3). Mutations were concentrated in the third codon positions (90.00%) and 96.66% were synonymous. The only exception was the gene *NAD2* where 3 non-synonymous substitutions led to amino acid exchange at positions 97, 319 and 434, being one a substitution of isoleucine by valine (both nonpolar amino acids) in the functional domain Oxidored g1 (http://www.uniprot.org/uniprot/Q9B8D2). As revealed by the alignments, *COB* and Cytochrome *c* oxydase subunit *1* (*COX1*) are the most variable genes among strains. These genes are the only intron containing genes in *C. albicans* mtDNA (*COB* has 3 exons and 2 introns while *COX1* has 5 exons and 4 introns) according to th annotation based on the highly similar sequences of *C. parapsilosis* mtDNA (available in the *Candida* genome database website). The 2,811 bp sequence of *COB* has 73 variable sites (43 transitions, 29 transversions and 1 deletion), located mainly in its introns (94.52%), with a frequency of nucleotide change (substitutions plus indels or “gaps”) of 2.59%. *COX 1,* which has 6,155 bp, has 59 variable sites (35 transitions, 20 transversions and 4 insertions) also concentrated in introns (69.49%), with a frequency of nucleotide change of 0.95%. Genes coding for proteins ATP8, ATP9 and all of the 30 genes coding for tRNAs did not show any mutation on both strains in comparison to SC5314.

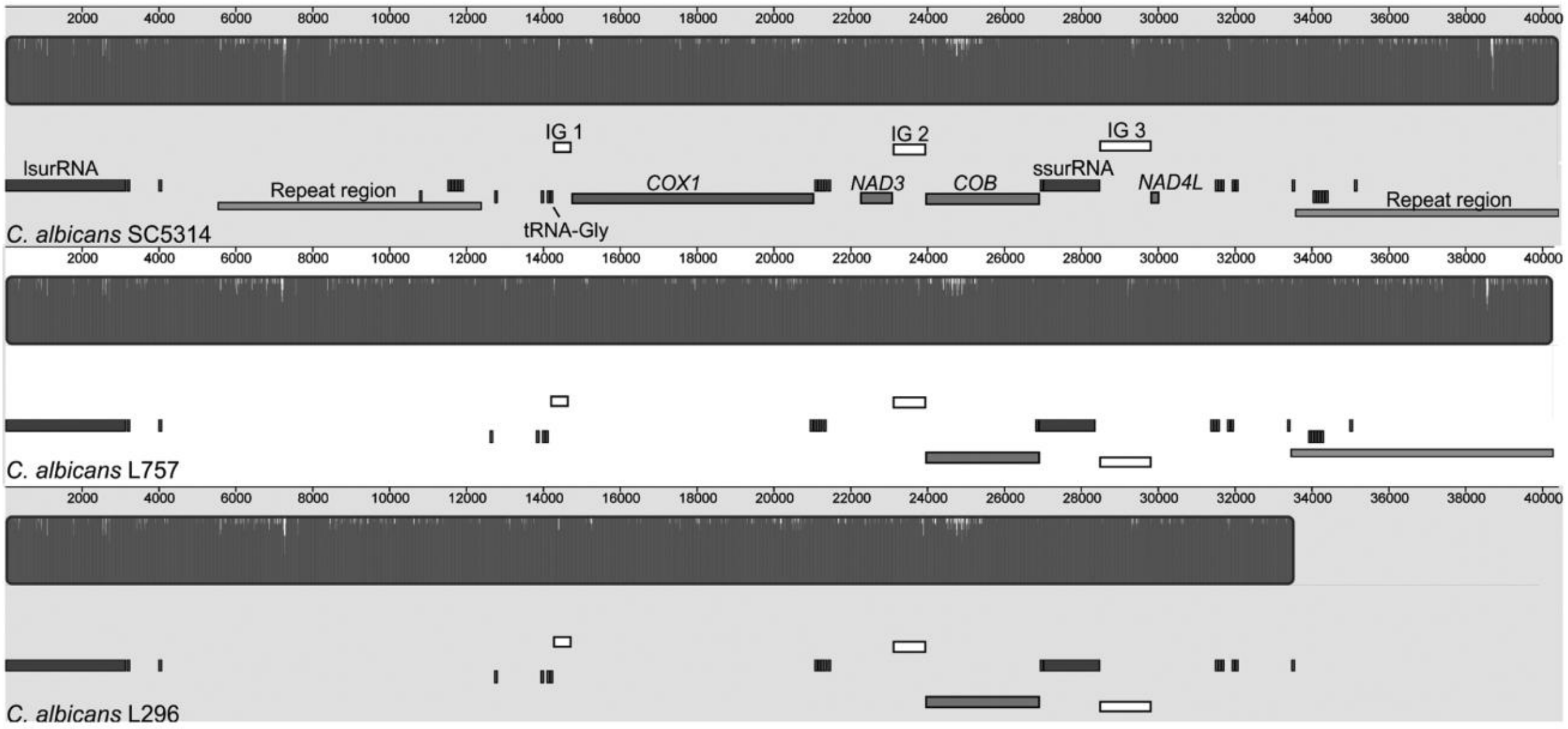

Fig.1. Alignment of *C. albicans* mitochondrial genomes of strains SC5314, L296 and L757 by the progressive mauve algorithm. Map positions of the three mitochondrial intergenic regions characterized in this study IG1 (tRNA-Gly/*COX1*), IG2 (*NAD3/COB*) and IG3 (ssurRNA/*NAD4L*) in white boxes. Grey boxes indicate coding genes and rRNA genes. Small bars indicate tRNA genes. Numbered scale bars indicate distance in base pairs and grey vertical bars just below the scale bars indicate the sequence similarity.

The remaining 142 polymorphic sites observed (0.42%) were located in intergenic regions. The rate of nucleotide substitution ranged between 4.35% for the 23 bp region between genes lsurRNA/*tRNA-Ala* and 0.8% in the 4,405 bp region flanked by *NAD1*/*COX3a* (Table 4). The nucleotide variation was higher in intergenic spacers than in mitochondrial genes because only 26.66% had the frequency of nucleotide substitution above 1%, while 56.25% of intergenic regions exhibited frequency of nucleotide substitution above 1%. Among non-coding mitochondrial regions analyzed, three intergenic sequences (tRNA-Gly/*COX1*, *NAD3*/*COB* and ssurRNA/*NAD4L*) were further investigated. These regions, named IG1, IG2 and IG3 respectively, were selected for PCR amplification and sequencing from additional *C. albicans* strains to investigate their potential as a tool for strain differentiation because of appropriate sizes for amplification (519, 758 and 1086 bp respectively), straightforward sequencing (not located in repeat regions) and high frequency of nucleotide substitution (above 1%).

**Table 3.** Mutations in mitochondrial genes of *C. albicans* clinical isolates L296 e L757 relative to the strain SC5314. * Genes in one of the two repeat regions present in *C. albicans* mtDNA. Ts=Transitions, Tv=Transversions, Del=Deletions, Ins=Insertions, Ns=Non-synonymous substitution and bp=base pairs.

| Gene | Size (bp) | Variable sites | Mutation Frequency (%) | Types of mutation (Numbers observed) | | | | |
|---|---|---|---|---|---|---|---|---|
| | | | | Ts | Tv | Del | Ins | Ns |
| lsurRNA | 3130 | 29 | 0.93 | 13 | 15 | 1 | - | - |
| *COX2* | 788 | 10 | 1.27 | 7 | 3 | - | - | - |
| *NAD6* | 440 | 3 | 0.68 | 1 | 2 | - | - | - |
| *NAD1* | 953 | 3 | 0.31 | 1 | 2 | - | - | - |
| **COX3a* | 809 | 6 | 0.74 | 4 | 2 | - | - | - |
| *ATP6* | 740 | 8 | 1.08 | 6 | 2 | - | - | - |
| *NAD2* | 1427 | 13 | 0.91 | 7 | 6 | - | - | 3 |
| *NAD3* | 389 | 3 | 0.77 | 2 | 1 | - | - | - |
| ssurRNA | 1461 | 1 | 0.07 | 1 | - | - | - | - |
| *NAD4L* | 254 | 1 | 0.39 | - | 1 | - | - | - |
| *NAD5* | 1658 | 9 | 0.54 | 6 | 3 | - | - | - |
| *NAD4* | 1394 | 6 | 0.43 | 5 | 1 | - | - | - |
| **COX3b* | 809 | 6 | 0.74 | 4 | 2 | - | - | - |
| *COB* | 2811 | 73 | 2.59 | 43 | 29 | 1 | - | - |
| *COX1* | 6155 | 59 | 0.95 | 35 | 20 | - | 4 | - |
| Total | | | | 135 | 89 | 2 | 4 | 3 |

**Table 4**. Mutations in mitochondrial intergenic regions of *C. albicans* clinical isolates L296 e L757 relative to the strain SC5314. * Intergenic regions in one of the two repeat regions present in *C. albicans* mtDNA. Ts=Transitions, Tv=Transversions, Del=Deletions, Ins=Insertions and bp=base pairs. Bold face indicates IG1 (tRNA-Gly*/COX1*), IG2 (*NAD3/COB*) and IG3 (ssurRNA*/NAD4L*).

| Mitochondrial intergenic region | Size (bp) | Variable sites | Mutation Frequency (%) | Types of mutation (Numbers observed) | | | |
|---|---|---|---|---|---|---|---|
| | | | | Ts | Tv | Del | Ins |
| lsurRNA/tRNA-Ala | 23 | 1 | 4.35 | - | - | - | 1 |
| **NAD1/COX3a* | 4405 | 13 | 0.29 | 5 | 8 | - | - |
| **COX3a*/tRNA-Lys | 39 | 1 | 2.56 | - | 1 | - | - |
| *tRNA-Lys/tRNA-Leu | 705 | 8 | 1.13 | 2 | 6 | - | - |
| *tRNA-Glu/*ATP9* | 542 | 2 | 0.37 | - | 2 | - | - |
| *ATP6/ATP8* | 124 | 1 | 0.80 | 1 | - | - | - |
| **tRNA-Gly/*COX1*** | 519 | 5 | 0.96 | - | 4 | 1 | - |
| *COX1*/tRNA-Arg | 139 | 1 | 0.72 | - | 1 | - | - |
| ***NAD3/COB*** | 758 | 10 | 1.19 | 5 | 5 | - | - |
| *COB*/tRNA-Met | 126 | 1 | 0.80 | 1 | - | - | - |
| **ssurRNA/*NAD4L*** | 1086 | 13 | 1.10 | 3 | 9 | 1 | - |
| tRNA-Ser/tRNA-Ser | 195 | 3 | 1.54 | - | 3 | - | - |
| *tRNA-Met/tRNA-Glu | 476 | 3 | 0.63 | - | 3 | - | - |
| *tRNA-Leu/tRNA-Lys | 675 | 8 | 1.18 | 2 | 6 | - | - |
| *tRNA-Lys/*COX3b* | 39 | 1 | 2.56 | - | 1 | - | - |
| **COX3b*/lsurRNA | 4400 | 71 | 1.61 | 37 | 32 | 1 | 1 |
| Total | | | | 56 | 81 | 3 | 2 |

### 3.2. Variability of mitochondrial intergenic regions between *C. albicans* strains.

The three mitochondrial intergenic regions selected IG1, IG2 and IG3 were sequenced in other 16 clinical isolates and 2 standard ATCC strains from different locations (Table 1): United States (2), Brazil (9 – São Paulo, Paraná and Bahia), Ecuador (2), Argentina (2), Venezuela (1),

Colombia (1) and one from an unknown location, from patients with hematogenic infections. Amplicons were obtained using the Elongase Enzyme Mix (Invitrogen) because of the 3'-5' exonuclease activity, which provides higher fidelity in polymerization than common Taq polymerase. PCR products were cloned prior to sequencing and the polymorphisms were confirmed by sequencing of independent PCR products to exclude artifacts of the amplification reaction and heteroplasmic effects. Identical results were obtained by direct sequencing of amplicons (data not shown).

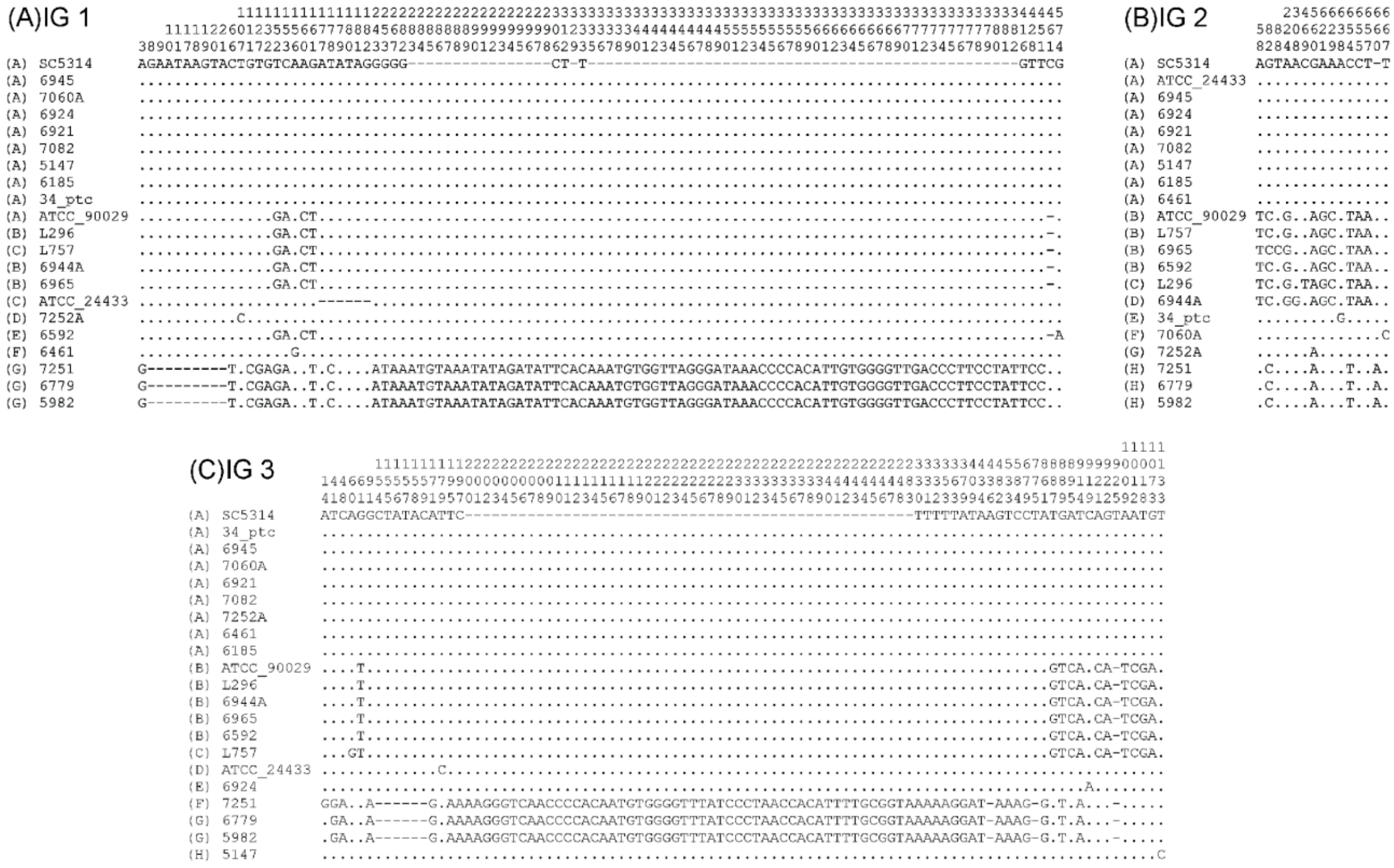

Fig. 2. Polymorphic sites in the mitochondrial intergenic regions IG1 (tRNA-Gly/*COX1*) **(A)**, IG2 (*NAD3/COB*) **(B)**, and (ssurRNA/*NAD4L*) **(C)** of 21 *C. albicans* clinical isolates and reference strains. Haplotypes are specified in parenthesis (A to H). Dots indicate nucleotides identical to the first sequence in the alignment and hyphens indicate indels (gaps).

The sequences obtained were aligned with their corresponding sequences of strains SC5314, L296 and L757, revealing a great variability in size and nucleotide sequence. The frequency of nucleotide changes (substitutions plus indels) were 19.84%, 1.98% and 8.65% for IG1, IG2 and IG3 respectively.

The alignment of IG1 sequences revealed 22 nucleotide substitutions and 81 indels ("gaps") between strains, totaling 103 variable sites (Fig. 2A). These 21 strains were distributed in 7 haplotypes (A to G) and their sequences exhibited great variability in size, ranging from 513 to 575 bp, with up to 56 bp difference relative to strain SC5314. The intergenic regions of the 2 Argentinean strains (5982 and 6779) and 1 from Brazil (7251) had identical nucleotide sequences between each other and were the most variable respective to other strains, especially because of two great indel segments of 16 and 48 bp (Fig. 2A positions 283 and 335). The IG2 alignment reveals 14 substitutions and 1 indel. These strain sequences were distributed in 8 haplotypes (A to H) (Fig. 2B). Alignment of IG3 sequences showed considerable variability between clinical isolates. 35 substitutions and 59 indel sites were observed, resulting in 94 variable sites distributed in 8 haplotypes (A to H) (Fig. 2C). Sequence sizes diverged up to 41 bp relative to strain SC5314, ranging from 1,085 to 1,127bp. The Argentinean strains (5982 and 6779) and strain 7251 from Brazil also have identical nucleotide sequences and two exclusive indels of 6bp and 49bp (Fig. 2C positions 154 and 200).

Modeltest (Posada and Crandall, 1998) was used to estimate the best fitting substitution model for the intergenic regions aligned. The models selected were TIM1+G, TPM3uf+I and F81 for IG1, IG2 and IG3, respectively. Bayesian trees were inferred for each sequence alignment using the corresponding substitution models. Tree topologies indicated that the isolates were distributed in three groups with posterior probabilities values above 60% (Fig. 3). Group 1 is formed mainly by isolates with the same or very similar haplotype as the reference strain SC5314, with 7 isolates from Brazil, 2 from United States, 1 from Venezuela, 1 from Colombia and 1 from Ecuador. Groups 2 and 3 are formed by strains that present more divergent sequences when compared to strain SC5314. Group 2 is formed by 4 samples from Brazil, 1 from United States and 1 from Ecuador while group 3 is formed mainly by the Argentinean strains (5982 and 6779), with the exception of one strain from Brazil (7251). These three strains were the most divergent and had exclusive indels segments of up to 49 bp relative to other strains. We assume that these indel segments, as well as the other exclusive polymorphisms present in these strains, might be geographically related and that migration could explain the presence of the Brazilian strain (7251) in this group. We did not detect any obvious clustering of strains in clades directly related to their geographic isolation, except for strains 5982 and 6779 from Argentina that grouped together forming group 3 along with the strain 7251 from Paraná, Brazil. To confirm the topology, phylogenetic trees were generated by Neighbor-joining and the relatedness of these strains was supported (data not shown).

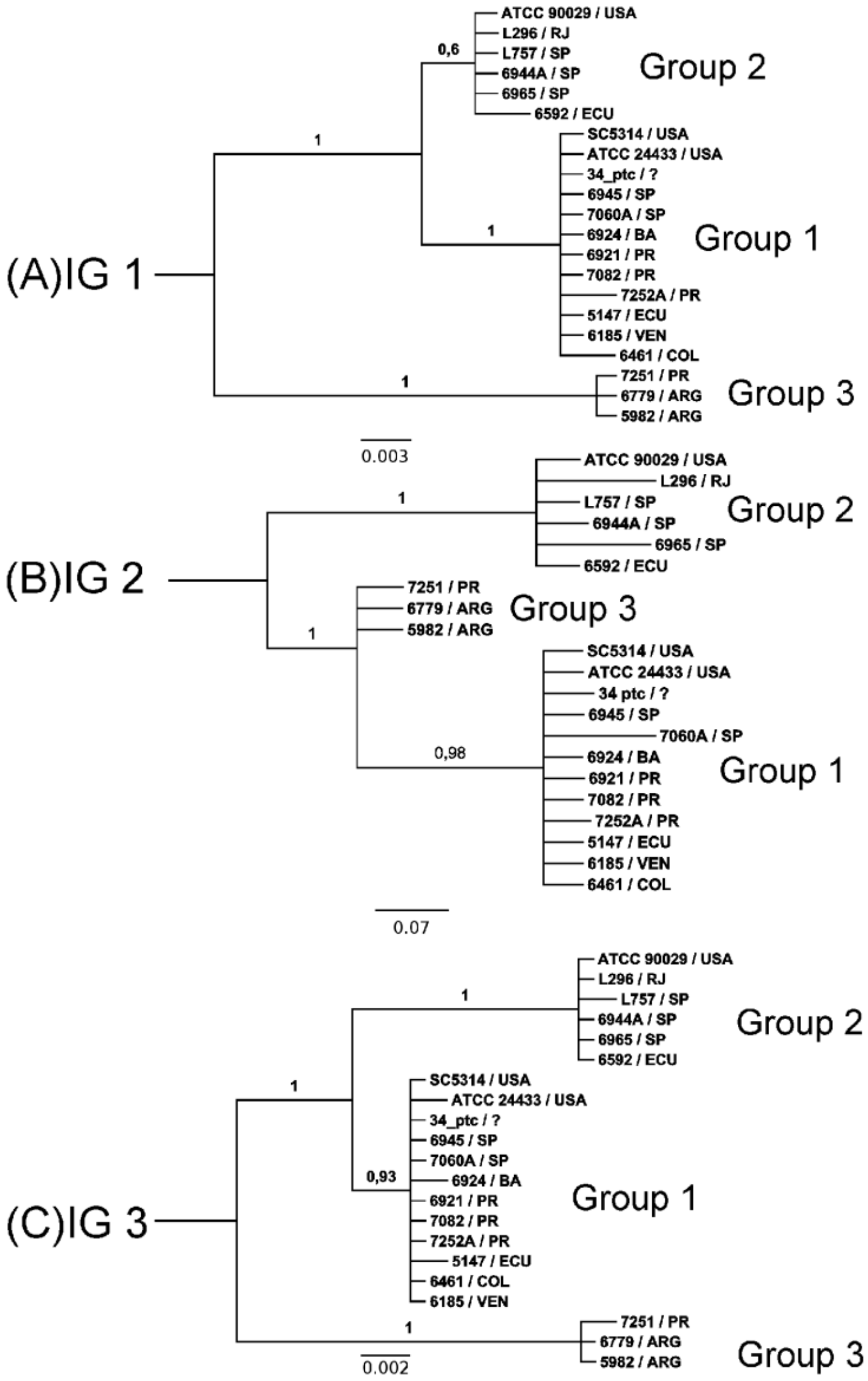

Fig. 3. Bayesian phylogenetic trees inferred from the nucleotide sequences of the mitochondrial intergenic regions IG1 (tRNA-Gly/*COX1*) **(A)**, IG2 (*NAD3/COB*) **(B)** and IG3 (ssurRNA/*NAD4L*) **(C)** of 21 *C. albicans* clinical isolates and reference strains. The topology of trees revealed the existence of three distinctive groups with posterior probabilities (numbers above branches) above 0.6. Scale bars indicate the number of substitutions per sequence position. Trees are depicted as midpoint rooted. The geographical origin of isolates is indicated just besides their identification codes: USA=United States, ARG=Argentina, COL=Colombia, ECU=Ecuador, VEN=Venezuela and BA, PR, RJ, SP are the Brazilian states of Bahia, Paraná, Rio de Janeiro and São Paulo, respectively.

To determine whether the most variable segments of the nucleotide sequences from the intergenic regions were not under positive selection, as expected for non-coding regions, it was performed the Tajima's D (Tajima, 1989) and Fu and Li's (Fu and Li, 1993) D* and F* tests of neutrality. The values obtained were not significantly different from zero, which indicates no deviation from the neutral model of evolution, and that the three intergenic regions are probably not under selective pressure (Table 5). These data indicate that the mutations in these intergenic regions IG1, IG2 and IG3 are not affected by natural selection and that estimates of distances are expected to follow the orthologous steps in the evolutionary pattern of these strains and will not underestimate the real number of events, which is a problem in sequences subjected to strong negative selection.

**Table 5.** Values for Tajima's D and Fu and Li's D* and F* obtained as an estimation from deviation of neutral evolution for the variable sites present at the mitochondrial intergenic regions analyzed in 21 *C. albicans* strains. (p>0.10 i.e. not significant).

| Intergenic region | Tajima (D) | Fu & Li (D*) | Fu & Li (F*) |
|---|---|---|---|
| (IG1) tRNA-Gly/*COX1* | - 0.00041 | 0.85637 | 0.7002 |
| (IG2) *NAD3/COB* | 0.65858 | - 0.33631 | - 0.05135 |
| (IG3) ssurRNA/*NAD4L* | 0.20895 | 0.86175 | 0.77592 |

Because of the great DNA exchange between the nucleus and mitochondria, we tested whether intergenic regions IG1, IG2 and IG3 could have been recently transferred to the nucleus. DNA was purified from isolated nuclei of *C. albicans* strain 5982 (Argentinean). Mitochondrial intergenic regions sequences from this strain have insertions of up to 49 bp that could result from unspecific amplification from nuclear DNA. To confirm if the primers designed specifically amplified the corresponding intergenic regions and that their localization was exclusively mitochondrial, the nuclear gene *ACT1* and the mitochondrial *COX2* were used as positive and negative controls, respectively. Amplification of *ACT1* was positive, confirming the presence of nuclear DNA in the sample, and no amplification of *COX2* and IG1, IG2 and IG3 was observed, indicating that the intergenic regions of interest are exclusively amplified from mitochondria and are not artifacts produced from nuclear sequence amplification (Fig. S1, Supplementary material). Additionally, we performed nucleotide BLAST using these mitochondrial intergenic regions as queries and found no matching outside the mitochondrial genome (data not shown).

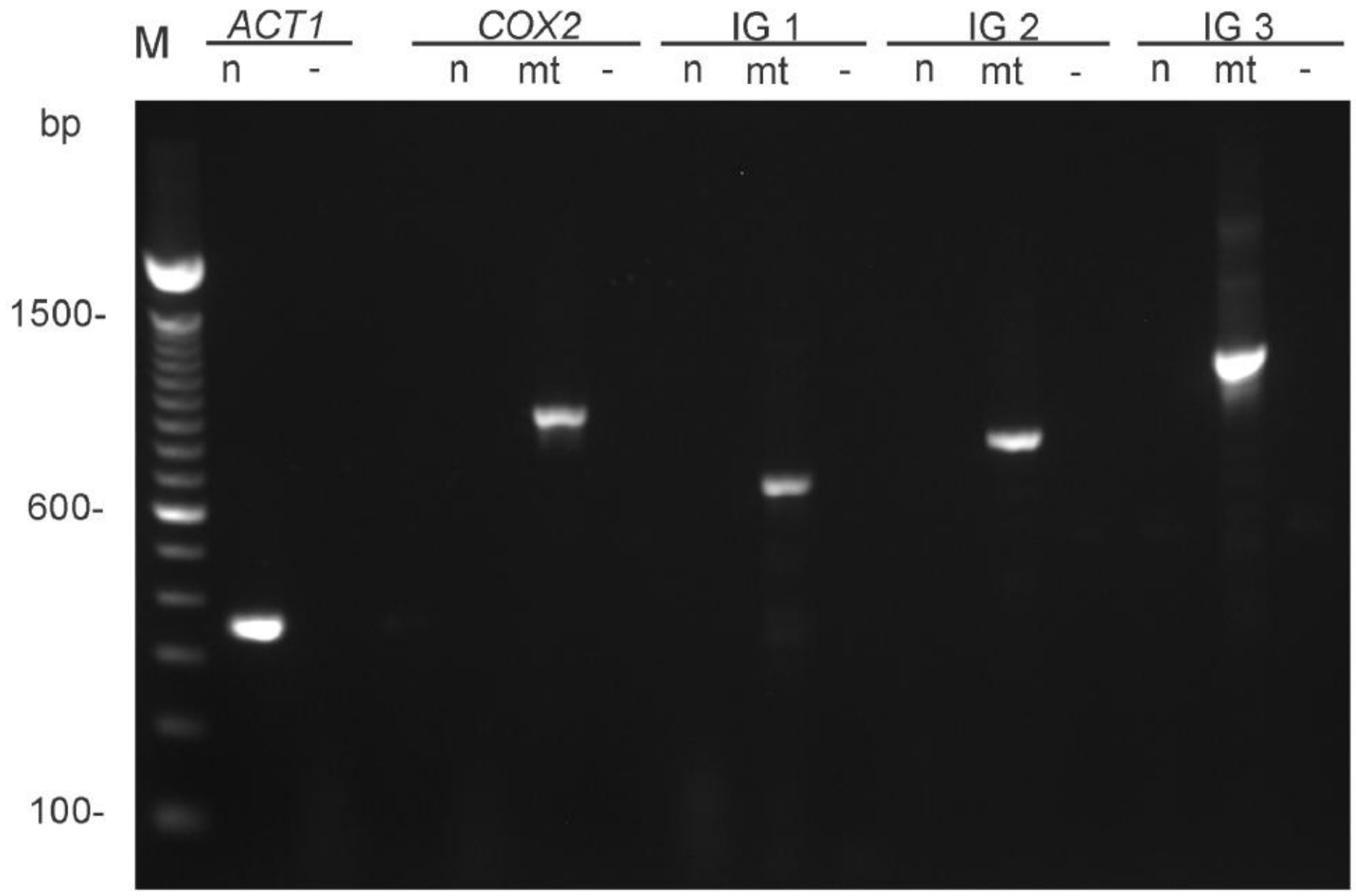

Fig. S1. Amplification reactions of *ACT1, COX2* and mitochondrial intergenic regions tRNA-Gly/*COX1*, *NAD3*/*COB* and ssurRNA/*NAD4L* in nuclear DNA. (n) nuclear DNA; (mt) mitochondrial DNA; (-) negative control (without DNA). Amplification of *ACT1* and *COX2* was used as positive and negative control respectively for nuclear DNA. MtDNA was used as a positive control for amplification of the mitochondrial gene *COX2* and the intergenic regions.

### 3.3. Intraspecific sequence variability of *COB.*

As revealed by the comparative sequence analysis of the whole mitochondrial genome of *C. albicans* strains L296, L757 and SC5314, *COB* was the gene with the higher number of mutations. To compare *COB* variability with the intergenic regions, we have amplified and sequenced this gene in the group 3 strains (the Argentineans 5982 and 6779 and the Brazilian 7251). Sequence alignment of *COB* along with the sequences from strains L296, L757 and SC5314 revealed 79 variable sites (78 substitutions, 1 deletion), and 3 haplotypes (Fig. 4). 93.67% of the mutations were located in introns and those located in exons were in the third codon positions producing only synonymous substitutions. Group 3 strains could not be differentiated from each other, and although they showed 6 exclusive polymorphisms, great part of these mutations were shared by the 5 strains (87.34%) relatively to SC5314. The frequency of nucleotide change (substitutions plus indels) was 2.81%, while the observed for intergenic regions IG1 and IG3 in these same strains were 18.49% and 8.38%, respectively. Overall pairwise mean distance of *COB* alignment was 1% while IG1, IG3 and IG2 was 2.3%, 1.6%, and 0.6% respectively.

```
                                                                       11111111111111111111111112
           1122455555666666667777777777777788888888888899999999900000000000012223333444881
           4911812344011255890134555667788802355556799900133457901112333466900563567002048
           4509713012935128995191458490278962923482968949105804771671457914613575122470248
(A) SC5314 taatACTCCTCAAAAGCTTAGCTTGGCCTTTAGAAGAACAACCTCATAAAACTAGATATAATTATAAATTAAGGTGATT
(B) L757   aggcGTCATCTGGGGAAACGCTACTATTACAGACTA.C.GGTACAGACGTT..GTTAGCCGGAGCGTTC.CTATCAG-.
(B) L296   aggcGTCATCTGGGGAAACGCTACTATTACAGACTA.C.GGTACAGACGTT..GTTAGCCGGAGCGTTC.CTATCAG-.
(C) 5982   .ggcGTCATCTGGGGAAA.GCTACTATTACAGACTAGCTGGTACAGACGTTTCGTTAGCCGGAGCGTTCACTATCA..c
(C) 7251   .ggcGTCATCTGGGGAAA.GCTACTATTACAGACTAGCTGGTACAGACGTTTCGTTAGCCGGAGCGTTCACTATCA..c
(C) 6779   .ggcGTCATCTGGGGAAA.GCTACTATTACAGACTAGCTGGTACAGACGTTTCGTTAGCCGGAGCGTTCACTATCA..c
```

Fig. 4. Polymorphic nucleotide sites in *COB* sequences between *C. albicans* group 3 strains (5982, 7251, 6779) and L296, L757 and SC5314. Uppercase letters indicate variable sites localized in introns. Haplotypes are specified in parenthesis (A to C). Dots indicate nucleotides identical to the first sequence in the alignment and hyphens indicate indels (gaps).

### 3.4. Comparison of the nuclear marker rDNA ITS and mitochondrial intergenic regions.

The rDNA ITS is a widely non-coding nuclear marker used in *Candida* species discrimination. We have sequenced and compared the mitochondrial IG1, IG2 and IG3 sequences variability to the ITS sequences obtained for the same 21 *C. albicans* strains. The overall pairwise mean distance observed for ITS1 and ITS2 was 0.2 and 0.5% respectively, while the average measured distances for IG1, IG2 and IG3 were 1.2%, 0.6% and 0.9% respectively, excluding gaps. These data indicate that the mitochondrial intergenic regions are significantly more informative, having at least 6 times the potential for strain discrimination than the nuclear non-coding marker ITS1. The divergence within IG1, IG2 and IG3 is in fact even greater if considered gaps (indel information). While these strains differed only in 1 site for ITS1 and 3 sites for ITS2, the IG1, IG2 and IG3 alignments revealed 103, 15 and 94 variable sites respectively, which conclusively show the unparalleled informative potential of these regions for strain discrimination. In Fig. 5 the phylogenetic tree inferred from the nuclear sequence rDNA ITS showed the same basal dichotomy as compared to mtDNA intergenic regions (Fig. 3) but because ITS has a smaller number of substitutions (smaller evolutionary rate), group 3 readily identified with mtDNA-IG sequences, is not observed in ITS trees. In other words the mtDNA IG shows an equivalent pattern but the polytomies (under 50% majority rule) indicate that ITS has less resolution than mtDNA-IG sequences. Therefore, the concern that nuclear and mitochondrial sequences, because of differences in ploidy, cell location and segregation could be "telling" different evolutionary "stories" is here not supported. Our data and analysis suggest that in the case of sequences here compared, the nuclear and mitochondrial sequences tell the same "story" although the mitochondrial "tells it" in more detail. The ITS rDNA Bayesian tree (Fig. 5) was inferred using the concatenated alignment of ITS1 and ITS2 sequences of 21 *C. albicans* isolates and the K80 transition matrix (Kimura, 1980) as indicated by model fit analysis implemented in program Modeltest (Posada and Crandall, 1998).

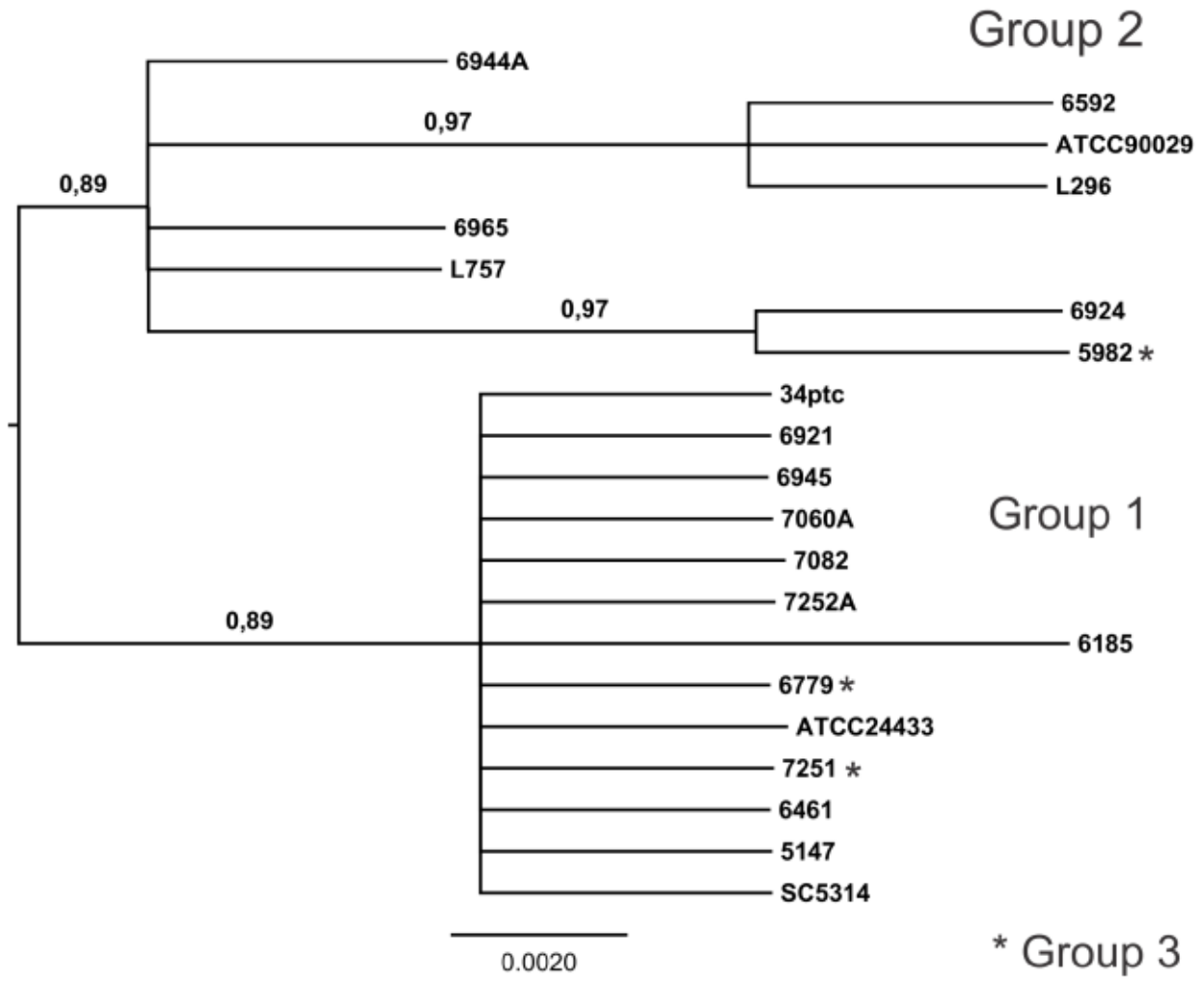

Fig. 5. Bayesian phylogenetic tree of rDNA ITS of *Candida albicans* isolates and strains. The Bayesian phylogenetic tree was inferred from the concatenated alignment of ITS1 and ITS2 nucleotide sequences of 21 *C. albicans* strains. The phylogeny is depicted as midpoint rooted tree. The scale bar indicate the number of substitutions per sequence position.

## 4. Discussion

Molecular typing methods are essential in epidemiology of *C. albicans.* These methods can also be useful in identifying the contamination source of outbreaks in the hospital environment, by differentiating strains according to microvariation in their genomes. MtDNA is more prone to revealing microvariability than commonly used nuclear targets, because of its higher mutational load and evolutionary rate (Clark-Walker, 1991). In this study, we have sequenced and compared two complete mitochondrial genomes of *C. albicans* strains with the reference SC5314, to investigate mtDNA variation in *C. albicans* and identify intraspecific hypervariable sites. We identified three intergenic regions, with great variability suitable for amplification and sequencing and further investigated their nucleotide diversity, phylogenetic pattern and modes of natural selection of mtDNA in *C. albicans* clinical isolates.

Comparative sequence analysis of the mtDNA of strains L296, L757 relative to SC5314, indicated mutation hot spots in the mtDNA, as revealed by analysis of mitochondrial gene sequences in phylogenetic related animals**,** such as humans and primates (Galtier et al. 2006). With exception of three non-synonymous changes in *NAD2*, the majority of mutations in the coding regions were synonymous and located in the third codon positions. Because mutations in these sites do not change the encoded protein, in theory, there is no effect on fitness (neutral) and therefore should reflect the evolutionary history of these strains and not adaptive changes (Gerber et al. 2001).

Approximately 36% (14,607 bp) of *C. albicans* mitochondrial genome comprises intergenic regions, most of them having few base pairs (up to 200 bp) or being located in one of the two repetitive portions of the mitochondrial genome, which makes sequencing very difficult. However, three of them, flanked by genes tRNA-Gly*/COX1*, *NAD3*/*COB* and ssurRNA*/NAD4L,* designated IG1, IG2 and IG3 respectively, showed potential for use in populational and molecular typing studies of *C. albicans*. Comparison of these regions in 21 *C. albicans* isolates (clinical and reference) showed a high frequency of nucleotide substitution (19.84%, 1.98% and 8.65% for IG1, IG2 and IG3 respectively), a value higher than most of the observed in the mitochondrial genes evaluated in this study. *COB* and *COX1*, were the most variable genes in the whole mtDNA sequence analysis, but still showed values lower than intergenic regions IG1 and IG3.

*COB* variability had already been used in inter and intraspecific molecular typing studies, including yeasts, such as genera *Candida* and *Trichosporon* (Biswas et al. 2001; Biswas et al. 2005; Yokoyama et al. 2000). Sequence comparison of *COB* between group 3 strains (5982, 6779 and 7251) with L296, L757 and SC5314, although having a high number of variable sites (79), including 6 exclusive for the group 3 strains, were not able to differentiate each sequence within group 3 since they shared the same polymorphisms. Intergenic regions IG1 and IG3 beyond enabling the discrimination of group 3 strains also showed a frequency of nucleotide substitution almost 6 and 3 times higher than *COB,* respectively*.* Biswas et al. (2001), typing of 32 *C. albicans* strains, found only three variable sites in a 396 bp segment, which corresponds to a frequency of nucleotide substitution of only 0.76%. Despite the high variation, intergenic regions are smaller (519 to 1086bp), easier to amplify and require fewer primers for full length sequencing.

Mutations in non-coding regions occur more frequently than synonymous changes in coding sequences and are among the most common evolutionary changes at the molecular level (Kimura, 1983). This higher number of mutations leads to faster evolutionary rates when compared to nuclear and mitochondrial genes, which makes them a good tool for studying closely related isolates (Watanabe et al. 2005). Ghikas et al. (2010) evaluated the potential of variations in the mitochondrial intergenic regions in intraspecific discrimination of the entomopathogenic fungus *Beauveria bassiana*. Analysis of the nucleotide sequences between genes *NAD3*/*ATP9* and *ATP6*/ssurRNA showed that their sizes were extremely variable among strains (73 and 200 bp difference respectively) and that, due to the large variability, these mitochondrial intergenic sequences allowed a better differentiation of strains than the sequence of the widely used nuclear marker rDNA ITS1-5.8S-ITS2. Furthermore, the authors also showed that although phylogenetic trees using the two separate data grouped the strains in similar clades, trees using concatenated ITS1 and mitochondrial intergenic regions data, resulted in subdivision of the major clade into seven distinct subgroups with some geographic association. In our analysis, the non-coding nuclear marker rDNA ITS also showed reduced number of polymorphisms among *C. albicans* strains than the mitochondrial non-coding IG1, IG2 and IG3. While ITS1 sequences showed only 1 polymorphic site, the mitochondrial intergenic regions showed up to 103 variations which indicate a 6 times greater variability among intraspecific isolates. In addition, sequences of intergenic regions allowed discrimination of these strains in three groups, including geographic differentiation of the Argentinean strains, while rDNA ITS could not (Fig. 5).

Although *Candida* spp*.* are human opportunistic pathogens with worldwide distribution, the existence of certain differences between isolates from different geographical locations is still expected and tend to increase with migration. This is possibly due to the action of independent evolutionary events in each strain, in separate areas and/or the existence of local reservoirs (non migratory) that are able to maintain certain strains associated with specific locations (Odds et al. 2007; Wrobel et al. 2008). Sanson and Briones (2000) studied *COX2* sequences of *C. glabrata* mtDNA and concluded that two polymorphic positions could be correlated with their geographical origin, discriminating strains from Brazil and United States. Nucleotide sequences from mitochondrial intergenic regions here analyzed were able to differentiate the Argentinean strains from others. The fact that a clinical isolate from the state of Paraná (Brazil) presented the same haplotype as these can suggest that the exclusive variations in their sequences are geographically related and the presence of such strain in this group is due to a migration event, especially because Paraná State borders Argentina. We have no further information about the patient's origin and in which city the isolate was obtained. Commonly the source of the individual's infection is the fungus present in his own microbiota, so the geographic location where the sample was isolated may not literally represent their reservoir and place of origin. In our analysis, no geographic association could be made with the clades observed in the phylogenetic inferences, except for the Argentinean strains. Molecular typing studies using MLST in *C. albicans* isolates tend to cluster them according to their geographical location; however, when using larger databases, these geographic data often become diluted and is no longer possible to make this distinction, only some suggestions of geographical enrichment of related strains (Odds et al. 2007). For this reason, further analysis, with a greater number of isolates, are needed to address the use of these intergenic regions for de facto utility as typing marker.

Some disadvantages may arise from the use of mtDNA as a molecular marker. In yeast, mtDNA escapes into the nucleus in a remarkably high frequency, although the opposite is not so often (Thorsness and Fox, 1990). Some technical problems arising from its use may be a consequence of displacement and insertion of fragments of mtDNA into the nuclear DNA, which can still be amplified with conserved primers, complicating and confusing the sequence analysis (Zhang and Hewitt, 1996; Bensasson et al. 2001). In this study, we were not able to amplify the selected mitochondrial intergenic regions in nuclear DNA or identify any similarity of these sequences with nuclear counterparts, confirming that the intergenic regions amplified actually are located in the mitochondria (Fig. S1- Supplementary material).

The use of mtDNA in populational studies may also be discouraged by indications that mtDNA is not strictly neutral and may be subject to positive selection more often than it is believed (Ballard and Kreitman, 1995; Hurst and Jiggins, 2005) because of the constant interaction with nuclear proteins, including the formation of four of the five complexes involved in electron transport chain and the vital importance of ATP for cell function (Ballard and Rand, 2005). Accordingly, it is recommended that population studies using mtDNA include statistical tests of neutrality (Ballard and Kreitman, 1995). In our study, we tested whether the nucleotide sequences of variable intergenic regions were under the effect of selective pressure by the methods of Tajima and Fu and Li (Fu and Li, 1993; Tajima, 1989). There were no deviations from neutrality, indicating that the variations found in the nucleotide sequences are in accordance with the neutral model of evolution, enhancing the potential use of these regions in typing studies due to its unconstrained variability.

## 5. Conclusions

The three mitochondrial intergenic regions analyzed here are easily obtained by PCR, sequencing and do not generate data that are dependent on subjective interpretation. Moreover, with the primers designed, they are also successfully amplified with total genomic DNA isolation (Wach et al. 1994), which is much faster and simpler to perform than mtDNA extraction (data not shown). These intergenic regions also showed high variability, even higher than mitochondrial genes and the non-coding nuclear marker ITS and showed a few polymorphisms that may be geographic related. Further analysis, with a larger and variable number of samples, is required to investigate the full potential of these mutations to discriminate geographic variants of *C. albicans*. Nevertheless, our data show, for the first time that mitochondrial intergenic regions IG1, IG2 and IG3, which evolves under neutrality and have a high nucleotide variability, can be expected to contribute in molecular studies concerning *C. albicans* strains along with other well established methods, such as MLST.

## Acknowledgements

We thank Bruno Giordano and Paloma Hernandez for technical assistance in performing the bulk sequencing of the *C. albicans* L296 strain mitochondrial genome. TFB received a MSc fellowship from Conselho Nacional de Desenvolvimento Científico e Tecnológico (CNPq), Brazil and RCF received a postdoctoral fellowship from Fundação de Amparo à Pesquisa do Estado de São Paulo (FAPESP), Brazil. This work was supported by grants to MRSB from FAPESP, Brazil; CNPq, Brazil and the International Program of the Howard Hughes Medical Institute.